# Orbits and resonances of the regular moons of Neptune

[1]Marina Brozović, [2]Mark R. Showalter, [1]Robert A. Jacobson, [2]Robert S. French, [3]Jack J. Lissauer, [4]Imke de Pater

[1]Jet Propulsion Laboratory, California Institute of Technology,
Pasadena, CA 91109-8099, USA

[2]SETI Institute, Mountain View, CA 94043, USA

[3]Space Science & Astrobiology Division, MS 245-3, NASA Ames Research Center, Moffett Field, CA 94035, USA, orcid ID 0000-0001-6513-1659

[4]University of California Berkeley, Berkeley, CA 94720, USA

Running head: Satellites of Neptune

Correspondence should be addressed to:
Dr. Marina Brozović
4800 Oak Grove Drive, Mail Stop 301-121
Jet Propulsion Laboratory, California Institute of Technology
Pasadena, CA 91109-8099 USA
Tel 818-354-5197, Fax 818-393-7116
E-mail: Marina.Brozovic@jpl.nasa.gov

**Abstract**

We report integrated orbital fits for the inner regular moons of Neptune based on the most complete astrometric data set to date, with observations from Earth-based telescopes, Voyager 2, and the Hubble Space Telescope covering 1981-2016. We summarize the results in terms of state vectors, mean orbital elements, and orbital uncertainties. The estimated masses of the two innermost moons, Naiad and Thalassa, are $GM_{Naiad}$= 0.0080±0.0043 km$^3$ s$^{-2}$ and $GM_{Thalassa}$=0.0236±0.0064 km$^3$ s$^{-2}$, corresponding to densities of 0.80±0.48 g cm$^{-3}$ and 1.23±0.43 g cm$^{-3}$, respectively. Our analysis shows that Naiad and Thalassa are locked in an unusual type of orbital resonance. The resonant argument $73\dot{\lambda}_{Thalassa} - 69\dot{\lambda}_{Naiad} - 4\dot{\Omega}_{Naiad} \approx 0$ librates around 180° with an average amplitude of ~66° and a period of ~1.9 years for the nominal set of masses. This is the first fourth-order resonance discovered between the moons of the outer planets. More high precision astrometry is needed to better constrain the masses of Naiad and Thalassa, and consequently, the amplitude and the period of libration. We also report on a 13:11 near-resonance of Hippocamp and Proteus, which may lead to a mass estimate of Proteus provided that there are future observations of Hippocamp. Our fit yielded a value for Neptune's oblateness coefficient of $J_2$=3409.1±2.9 × 10$^{-6}$.

# 1. Introduction

The Neptune system consists of seven regular inner moons, Triton, Nereid, and five irregular outer moons. Naiad, Thalassa, Despina, Larissa, Galatea, and Proteus are the regular moons discovered by the Voyager 2 spacecraft during the 1989 flyby of Neptune (Smith et al., 1989; Owen et al., 1991). Showalter et al. (2013) reported on the HST discovery of the seventh regular moon – Hippocamp. This tiny moon, ~17 km in radius orbits between Larissa and Proteus. All regular moons have nearly circular prograde orbits and their semi-major axes span 48,000-118,000 km. The semi-major axis of the farthest moon, Proteus, is less than five Neptune radii from the planet's center. The entire closely-packed system would fit within one third of the Earth-Moon distance.

Triton, discovered in 1846 by William Lassell (Lassell, 1846), is one of the oddballs of the solar system because it has characteristics of both regular and irregular moons. Triton is very large, ~1,350 km in radius and orbits close to the planet in a nearly perfect circle as is expected for a regular satellite. However, the orbit of Triton is retrograde with an inclination of ~157° with respect to Neptune's equator, as is typical for an irregular satellite. The regular and irregular moons have dramatically different origins; the regular moons were formed "in-situ" while the irregular moons are widely regarded as captured objects (Harrington and Van Flandern, 1979; Goldreich et al., 1989). Current theories suggest that Triton is a captured moon after all, and that the original regular moons got destroyed in collisions as Triton destabilized the system (McCord, 1966; Goldreich et al., 1989; Banfield and Murray, 1992; Ćuk and Gladman, 2006). Nereid is a possible lone survivor of the original population that was scattered into a highly eccentric orbit during the capture and orbital circularization of Triton. Nereid is sometimes regarded an irregular moon although its periapse and inclination are not typical of irregular moons.

The regular moons we see today formed from the ring of a collisional debris. They remain strongly perturbed by massive Triton. Triton is a critical component of the Neptunian system as the system's invariable plane is defined by both the rotational angular momentum of the planet and by the orbital angular momentum of Triton. Our paper describes the orbital fits for the regular moons based on the latest set of astrometry. Owen et al. (1991) were the first to publish orbital fits for the six regular moons known at the time based on the Voyager 2 imaging data. The orbits were modeled as precessing ellipses inclined to their respective Laplace planes. Owen et al. (1991) estimated seven equinoctial elements plus three stochastic pointing offset angles to account for the unknowns in the Voyager 2 camera orientation. Jacobson and Owen (2004) extended this study by adding Earth-based and HST astrometry to the original Voyager 2 data set. Because the data were still relatively sparse, Jacobson and Owen (2004) continued to use an analytical model to fit the data. The new orbital fits improved estimates of the mean motions and also determined the Laplace planes' precession rates. Jacobson (2009) was the first to attempt an integrated orbit fit to the Proteus astrometry and also revised the pole model for Neptune.

We report on integrated orbital fits for all seven regular inner moons of Neptune. This is a significant improvement over the precessing ellipse fits because integrated orbits rigorously capture the dynamics of the system. Our fits benefit from an extensive set of the high-precision HST astrometry that was obtained during 2004−2016 (Showalter et al., 2019), including all astrometry for Hippocamp.

## 2. Methods

### 2.1 Observations

Tables 1 and 2 summarize the observations used in this analysis. The astrometric data cover the period from 1981–2016, with the most significant amount of data originating from the Voyager 2 spacecraft and HST. Voyager 2 imaged all regular satellites except Hippocamp between 1989 June 7 and 1989 August 24 (Table 1). The narrow-angle camera took pictures of the satellites against a stellar background, and the stars provided the pointing reference for the camera. The camera had ~10 μrad resolution per sample (Owen et al. 1991). The positions of the satellites were recorded in the form of samples and lines. We used the astrometry reprocessed with the US Naval Observatory CCD Astrograph Catalog, UCAC2 (Zacharias et al., 2004) as described in Jacobson and Owen (2004). We applied tighter data weights than the ones reported in Jacobson and Owen (2004) because we noticed that the normalized fit residuals were too small and that the data were underutilized. Our newly assigned data weights were based on the size of a satellite and the distance between the spacecraft and the satellite, with a minimum data weight of 0.5 sample, corresponding to ~5 μrad. For example, on 1989 June 7 the spacecraft distance from Proteus was ~1×$10^8$ km, and the assigned 1σ position measurement accuracy was ~540 km. On 1989 August 23 Proteus was ~2.7×$10^6$ km from the spacecraft and the 1σ accuracy was 80 km. Given that the radius of Proteus is ~210 km, these are both very precise measurements. Naiad position accuracy ranged from ~120 km from a distance of 2.5×$10^7$ km to ~20 km from a distance of ~1.9×$10^6$ km. Naiad is the second smallest inner moon with a radius of ~33 km (Karkoschka, 2003).

The Earth-based and HST observations are listed in Table 2. The table starts with the observation of Larissa from 1981 from the University of Arizona Catalina and Mount Lemmon telescopes. This measurement predates the 1989 Voyager 2 discovery of Larissa. The satellite location was assumed to be coincidental with the occulted star location. The astrometry was reduced with respect to the UCAC2 star catalog (Jacobson and Owen, 2004) and the uncertainty in the star position was ~200 mas.

The follow-up observations originated from several Earth-based telescopes, but the majority were still obtained by HST. Most of the astrometry was reported as relative separations between a satellite and the center of Neptune in the tangent plane. Sicardy et al. (1999) detected Galatea on a single day in 1998 from the Mauna Kea Observatory and Colas and Buil (1992) observed Proteus on a single day in 1991 from the European Southern Observatory at La Silla. Marchis et al. (2004) observed Despina, Galatea, Larissa, and Proteus from the Mauna Kea Observatory in 2002 and 2003. They also reported three points for Thalassa and three tentative detections of Naiad. Jacobson and Owen (2004) used these points in their fit, but with very loose weights of 225 mas for Naiad and 60 and 180 mas for Thalassa. We used two out of three points for Thalassa weighted at 200 mas, but we did not use Naiad measurements because their residuals were large. Showalter et al. (2019) also found that the Marchis et al. (2004) observations of Naiad cannot be fit, and they reported that Thalassa's measurements were 19° off from their predicted position. We were able to fit two Thalassa points from 2002 with RA residuals of -327 mas and 546 mas and Dec residuals of -213 mas and 321 mas, respectively. These residuals are within 3σ of their weights, so we kept them in the fit.

Dumas et al. (2002) and Pascu et al. (2004) reported the HST observations of Despina, Galatea, Larissa, and Proteus. Jacobson and Owen (2004) converted the original Dumas et al. (2002) measurements of the longitude offsets from the Voyager ephemerides RNG022 (Owen et al. 1991) to Neptune-relative position angles. Vieira Martins et al. (2004) reported more Proteus astrometry spanning 2000–2002. These were relative measurements of Proteus position with respect to Triton as opposed to Neptune. We kept the same weights for the Earth-based observations as the ones published in Jacobson and Owen (2004).

Showalter et al. (2019) published the latest set of the HST astrometry including the discovery and follow up observations of Hippocamp. These measurements were obtained during 2004–2016. The satellite positions were measured relative from the center of Neptune in the tangent plane, and we kept the original weights as reported in the paper.

## 2.2 Orbital model

Jacobson et al. (2004) used a precessing ellipse fit referred to the Laplace plane of each satellite. This simplified model accounts for the effects of Neptune's oblateness and the Triton torque, but it does not account for the mutual interactions of the satellites due to their masses. Furthermore, it does not account for the gravitational perturbations from Triton, planets other than Neptune, or the Sun. We used integrated orbits to fit the data, then precessing ellipses in order to summarize the results of integration in terms of the mean orbital elements. Our orbital fit is based on the Peters (1981) formulation of numerical integration of the satellites of the outer planets. The equations of motion are defined in Cartesian coordinates with the Neptune system barycenter at the origin and referenced to the International Celestial Reference Frame (ICRF). The equations include the gravitational effects of Neptune, the $J_2$ and $J_4$ zonal harmonics of its gravity field, and perturbations by Triton, Jupiter, Saturn, Uranus, and the Sun. The mass of the Sun is augmented with the masses of the terrestrial planets and the Moon. The values of the dynamical parameters are listed in Table 3. We used the trigonometric expansion for the pole of Neptune from Jacobson (2009) and the pole parameters were kept fixed in the integration. We used DE435 planetary ephemerides (Folkner, 2016) for the orbits of planets and NEP081 (Jacobson, 2009) for the orbit of Triton. DE435 is a minor update of DE430 (Folkner et al., 2014). Compared to DE430, DE435 has an improved orbital solution for Pluto as well as minor improvements and/or updates for the orbits of Mercury and Saturn (Bill Folkner, pers. comm.).

The integration utilizes variable size time steps and is based on the variable-order Gauss-Jackson method. We set the maximum velocity error of $10^{-11}$ km s$^{-1}$ in order to control the integration step size. The average integration step size was 200 s. We used the weighted least-squares method based on the Householder transformations (Lawson and Hanson, 2004; Tapley et al., 2004) to solve for the state vectors, the *GM*s of the satellites, and *J₂*. In addition to these dynamical parameters, we also fit for a systematic camera bias in the Voyager 2 data. A single pair of sample and line biases is applied to all Voyager 2 imaging data.

## 3. Results

### 3.1 The state vectors, *GMs*, and *J₂*

Our fitting process progressed in several stages as we investigated which dynamical parameters can be fit based on the available astrometry. We started with a model where all inner satellites were kept massless and the state vectors were the only fitted parameters. This was followed by a fit where only the satellites that showed some mass sensitivity (Naiad and Thalassa) had their *GM*s fitted, while the rest were still considered massless. Finally, we progressed to a full fit where we estimated state vectors, *GMs* for all satellites (assuming some *a priori* values), and *J₂*.

The state vectors and the uncertainties of the final fit (nep090) are listed in Table 4. Galatea, Despina, Larissa, and Proteus have many observations and long data arcs that constrain their state vectors down to several kilometers (1σ). The small satellites, Naiad, Thalassa, and Hippocamp, have 1σ position uncertainties of 10-20 km, which is still acceptable accuracy given the sparse data set.

Figures 1 and 2 show the Naiad and Thalassa nep090 fit residuals for the two most abundant data sets: the Voyager 2 astrometry and the HST astrometry from Showalter et al. (2019). Overall, the root-mean-square (rms) of the residuals show good correspondence between the fit and the data. For the Voyager 2 data, Naiad has a combined rms of the line and sample residuals of ~49 km and Thalassa has a combined rms of ~60 km. The HST astrometry of Naiad has a combined rms of the X and Y residuals of ~21 mas and Thalassa has a combined rms of ~16 mas. At Neptune's average distance from Earth, 1 arcsec is 21,000 km, which makes the HST astrometry of Naiad and Thalassa almost an order of magnitude less precise than the Voyager 2 data.

The last two columns in Tables 1 and 2 list the nep090 rms of the residuals for all satellites. We evaluate the data weights by dividing the rms of residuals by the astrometric uncertainties. If the normalized rms is less than 1, the data weights were too loose, and if it is higher than 1, the data weights were too tight. All of our fits show normalized rms of ~1, suggesting that the data weights adequately describe the data quality.

Long data arcs and high-precision astrometry are prerequisites for good quality integrated orbits and *GM* estimates. The inner satellites of Neptune have data arcs that are several decades long, but the data are relatively sparse, especially for Naiad and Thalassa. Table 5 shows a range of *GM*s assuming densities between 0.05-1.5 g cm$^{-3}$. This range of densities was constrained by Zhang and Hamilton (2007) based on a theoretical study of the resonant history of Proteus and Larissa. Naiad, Thalassa, and other inner moons likely formed in a similar way to Proteus and Larissa, so we assumed that the same range of densities would apply to these moons as well.

The fitting process revealed the data sensitivity to the masses of Naiad and Thalassa, but not to the masses of other moons. The first estimates of the densities of Naiad and Thalassa suggested a range between 0.8-1.2 g cm$^{-3}$, and we used this as an *a priori* for the *GMs* of the other satellites (Table 5). The *a priori* values were removed to estimate the final uncertainties of the converged fit.

The fitted *GMs* for Naiad and Thalassa are $GM_{Naiad}$= 0.0080 ± 0043 km$^3$ s$^{-2}$ and $GM_{Thalassa}$= 0.0236 ± 0.0064 km$^3$ s$^{-2}$. The improvements in the means and the rms of the residuals are subtle between the massless and with-mass (nep090) fits (red and black residuals in Figures 1 and 2 and Table 6), suggesting that the data have a border-line sensitivity. However, the fit converged toward non-zero masses without any constraints. We also note that the masses of Naiad and Thalassa did not change significantly between the fit that had the rest of the satellites massless and the nep090 fit that estimated the *GM*s of the other satellites with Table 5 *a priori* constraints.

The Voyager 2 astrometry is important for the Naiad and Thalassa mass determination because it significantly extends the data arc. The uncertainties increased by more than 60% after we removed the Voyager 2 data from the fit. We also investigated how the relative weights between Voyager 2 and Showalter et al. (2019) data affect the mass determination. We de-weighted Voyager 2 data by a factor of two and re-converged the fit, which led to ~0.5σ change in the mass of Naiad and ~0.03σ change in the mass of Thalassa. This shows that the formal uncertainties successfully account for the mass changes due to a slightly different choice of the data weights.

Figure 3 shows the in-orbit, radial, and out-of-plane differences for the massless and nep090 fits. The in-orbit differences for Naiad show a sinusoidal pattern with an amplitude of ~500-600 km and a period of ~2 years. The small upward slope is due to a change in the mean motion. The out-of-plane orbit comparison also shows a difference of several hundred km. Thalassa's in-orbit differences show the same sinusoidal pattern as for Naiad, except these are smaller, ~200-300 km in amplitude, and they are shifted in phase by 180°. Given that the HST astrometry precision for Naiad and Thalassa is ~10-20 mas or, at the average Neptune's distance, ~200-400 km, it can be expected that the orbital fit is sensitive to dynamics that shifts the orbits several hundred km. The correlated pattern of Naiad and Thalassa in-orbit differences strongly suggests that these two satellites are in a resonance (Section 3.3).

The fitted *GMs* for Naiad and Thalassa allow for a density estimate. If we adopt the sizes from Karkoschka (2003) the densities are $\rho_{Naiad}$= 0.80 ± 0.48 g cm$^{-3}$ and $\rho_{Thalassa}$= 1.23 ± 0.43 g cm$^{-3}$ (Table 5). The uncertainties were calculated from the errors on the *GMs* and the radii. The Naiad and Thalassa densities are consistent with each other within the error bars and are also consistent with the Zhang and Hamilton (2007) constraints. The *GM*s of other moons did not change much from the values set by a default density of 1 g cm$^{-3}$, confirming our assessment that the current data have no mass sensitivity. The formal uncertainties on the *GMs* of Despina and Proteus are only about a factor of two larger than their *a priori* values, raising a possibility that their masses will become measurable with longer data arcs and more high-precision astrometry.

Despina, Galatea, and Larissa are closely packed with their semimajor axes, *a*, of 52,500 km, 61,900 km, and 73,500 km, respectively. The largest inner moon, Proteus (*a*~117,600 km), is fairly close to both Hippocamp (*a*~105,300 km) and Larissa. We compared the massless and nep090 orbital solutions from 1990 to 2030, and the most significant differences are evident for the orbit of Hippocamp. Hippocamp's in-orbit difference grows to ~100 km due to the influence of Proteus. For comparison, Larissa's in-orbit differences are on the order of ~20 km which is non-detectable with the current astrometry. We conclude that a several decades long data arc on Hippocamp with a few mas precision could allow for an estimate of the mass of Proteus. The *GMs* for Despina, Galatea, and Larissa also may become available given the sub-mas astrometry precision.

We found that the current fit improved the uncertainty on $J_2$ by almost a factor of two with respect to Jacobson (2009), from $J_2$=3408.4±4.5 × $10^{-6}$ to $J_2$=3409.1±2.9 × $10^{-6}$. The data showed no sensitivity to the value of $J_4$, but we still used $J_4$= -33.4±2.9 × $10^{-6}$ from Jacobson (2009) as a "consider parameter". Consider parameters are not estimated in the fit, but their uncertainties inflate the overall fit statistics, and result in more conservative errors (Biermann, 1977). We also investigated sensitivity of the fit to Neptune's Love number, $k_{2N}$, but the state vectors changed by only a few hundred meters when we hardwired the $k_{2N}$ value to 0.41 (Burša, 1992). The formal error on $k_{2N}$ was orders of magnitude larger than the value, signaling that $k_{2N}$ is unconstrained by the data at hand. We examined the data sensitivity with respect to the *GM* of Triton, but our fit showed an order of magnitude less sensitivity than that found by Jacobson (2009). This is not surprising considering that the dominant data for computing Triton's *GM* originate from the Voyager 2 flyby, used in Jacobson (2009) fit. Even when we changed Jacobson (2009) *GM* of Triton by 2σ, from 1427.6 km$^3$ s$^{-2}$ to 1423.9 km$^3$ s$^{-2}$, and re-converged the fits, orbits changed at sub-km level for all but Proteus and Hippocamp. Their orbits changed by a few km in the in-orbit and out-of-plane directions from 1990-2020. The data at hand are not sensitive to this level of change.

## 3.2 Mean Orbital elements

We summarize results of the orbital integration in the form of planetrocentric mean orbital elements and angle rates (Table 7) as we did in our previous analyses (Jacobson, 2009; Brozović and Jacobson, 2009; Jacobson et al. 2012; Brozović and Jacobson, 2017). We used 900 years of integrated orbits to make a large set of state vectors that were fit by precessing ellipses. The initial equinoctial elements and rates were refined by the least-squares method until the fit showed no further improvement. The fitted ellipses capture the constant and secular behavior of the integrated orbits and the residuals are dominated by the periodic perturbations. In this definition of mean elements, we do not associate any uncertainties with them because that would only describe how well the precessing ellipse fits to the integration. Only the state vectors and the *GMs* obtained from the integrated orbits have accompanying uncertainties.

The equinoctial elements and angle rates that describe the fits are usually reported with respect to an invariable plane of the system. However, Triton perturbs the moons so that their orbits precess about their local Laplace planes. Each Laplace plane is inclined by $i_{Lap}$ to the invariable plane of the Neptune system. Furthermore, the satellites have the orbits inclined with respect to their local Laplace planes by "free inclinations", $i_{free}$. The elements and rates in Table 7 are defined with respect to the individual Laplace planes. The node of the orbit, $\Omega_{free}$, is measured from the intersection of the local Laplace plane and the invariable plane (Figure 7 in Zhang and Hamilton, 2007). The local Laplace planes maintain fixed orientations with respect to Neptune's equator and Triton's orbit which means that they precess together with the Neptune's equatorial plane and Triton's orbit about the invariable plane at rate $d\Omega_{Triton}/dt$.

Overall, the mean elements from Table 7 match the Jacobson and Owen (2004) elements. We were able to estimate both $i_{Lap}$ and $i_{free}$ while Jacobson and Owen (2004) used a calculated value of $i_{Lap}$. $i_{Lap}$ progressively increases from Naiad, $i_{Lap} \sim 0.5°$, to Proteus, $i_{Lap} \sim 1°$, reflecting stronger influence from Triton. Naiad has a free inclination of ~4.7°, which is significantly higher than any other inner satellite of Neptune.

All satellites have short orbital periods, ranging from ~0.3 days for Naiad to just over a day for Proteus. The orbits are also characterized by relatively short nodal and apsidal precession periods ranging from ~0.3 years for Naiad to ~13 years for Proteus. The satellites have fairly circular orbits with the highest eccentricity being ~0.0012 for Larissa. Zhang and Hamilton (2007) studied the Proteus-Larissa 2:1 resonant passage that is the most likely culprit for the eccentricity of Larissa. Even today, the mean motion rates in Table 7 show that Proteus and Larissa are not far from their 2:1 resonance.

### 3.3 Two body resonances

Zhang and Hamilton (2007, 2008) presented a detailed numerical investigation of the past resonances in the Neptune system, but our analysis uncovered a resonance that is currently acting on Naiad and Thalassa. We found that the mean longitude rates and the nodal rate of Naiad add up to almost zero in the expression for the fourth-order inclination resonance, $\varphi = 73\dot{\lambda}_{Thalassa} - 69\dot{\lambda}_{Naiad} - 4\dot{\Omega}_{Naiad} = 6.6 \times 10^{-6} \, °\, day^{-1}$, for $\dot{\lambda}_{Naiad} = 1222.8584171°\, day^{-1}$, $\dot{\lambda}_{Thalassa} = 1155.7585817°\, day^{-1}$, and $\dot{\Omega}_{Naiad} = -1.71358060°\, day^{-1}$. Note that the Laplace inclinations of Naiad and Thalassa differ by only one milidegree, hence we can consider that their elements are in the same reference frame. We list an unusual number of digits because these rates are fits to the integrated orbits as opposed to the data. If we add up the mean longitude angles from Table 7, we get 180.023° which confirms that the two satellites are in a resonance.

Figure 3 already hinted at the mutual perturbations, but now we found the specific resonant angle responsible for the libration of Naiad and Thalassa. Figure 4 directly displays the resonant angle $\varphi$ calculated from osculating elements based on integrated orbits (nep090). A Fourier fit revealed an average amplitude of ~ 66° and a period of ~ 1.9 years. These values change for a different set of masses. As an example, we hardwired the mass of Thalassa to 0.03 km$^3$ s$^{-2}$, or +1$\sigma$ from the nominal mass, and re-converged the fit. The resonant argument still librates around 180°, but now with an amplitude of ~ 46° and a period of ~ 1.7 years.

The latest addition to Neptune's system, Hippocamp (Showalter et al., 2019), has the smallest eccentricity, $e=1\times10^{-5}$, and the smallest free inclination, $i_{free}=0.0019°$, among all inner moons of Neptune. Hippocamp appears to be very close to the 13:11 mean motion resonance with Proteus. This is likely the reason why our fit shows some sensitivity to the mass of Proteus (Table 5). As Proteus is migrating outward due to tidal interactions with Neptune, growth in its semimajor axis of only several tens of km will bring it into resonance with Hippocamp. We use the equation from Murray and Dermott (2001) (Section 4.9) that describes radial rate for a tidal migration of a small satellite,

$$\dot{a} = \pm \frac{3k_{2N}}{Q_N} \left(\frac{R_N}{a}\right)^5 \left(\frac{m}{m_N}\right) na \ , \qquad (1)$$

in order to get a first-order estimate of when Hippocamp and Proteus will enter the resonance. We assume Neptune's Love number $k_{2N}= 0.41$ (Burša, 1992) and tidal dissipation factor $Q_N = 20,000$ (Figure 2 in Zhang and Hamilton, 2008). The semi-major axis, $a$, mean motion, $n$, and mass of Proteus are listed in Tables 5 and 6. The mass of Neptune, $m_N$, can be obtained from the system mass by subtracting the mass of Triton and the rest of the regular satellites (Tables 2 and 5). The radius of Neptune $R_N$ was rounded to 24,600 km. Equation 1 gives an estimate of ~18 million years for Proteus to migrate ~40 km outward, which is approximately the location of resonance with Hippocamp.

## 4. Discussion

### 4.1 Naiad-Thalassa fourth-order resonance

The higher-order resonances, or the ones that involve inclinations are rare among the planetary satellites (Murray and Dermott, 2001). Saturn's satellites Mimas and Tethys are in the second-order mean motion resonance that involves both of the nodes. The resonant argument is of the form, $\varphi = 4\lambda_{Tethys} - 2\lambda_{Mimas} - \Omega_{Tethys} - \Omega_{Mimas}$ (Greenberg, 1973; Murray and Dermott, 2001), and the conjunction of the satellites librates about the midpoint of the nodes with an amplitude of 43.6° and a period of 71.8 years. Cooper et al. (2008) reported the first-order mean motion resonance between Mimas and Anthe that involves nodes of both satellites and the longitude of pericenter, $\varpi$, of Anthe: $\varphi = 11\lambda_{Anthe} - 10\lambda_{Mimas} - \varpi_{Anthe} - \Omega_{Anthe} - \Omega_{Mimas}$. Cooper et al. (2008) also noted the 77:75 eccentricity-type near-resonance between Methone and Anthe.

Fourth-order mean motion resonances are known to exist in the main asteroid belt. For example, the 7:3 resonance with Jupiter is located between the Koronis and Eos families, and some short-lived asteroids reside in this gap (Yoshikawa, 1991; Tsiganis et al., 2003). Furthermore, there are some trans-Neptunian objects (TNOs) that are in 1:5 resonance with Neptune (Gladman et al., 2012; Alexandersen et al., 2016; Bannister et al., 2018). We did not find a discussion of whether the inclinations or eccentricities play a role in these resonances.

The Naiad and Thalassa 73:69 resonance is the first case of the fourth-order resonance in a population of the planetary satellites and the first resonance that only involves a single satellite's node. Naiad and Thalassa 73:69 commensurability has five possible inclination-only dependent resonant terms (Table B.16 in Murray and Dermott, 2001):

$$\langle \mathcal{R} \rangle_1 = f_1 \, i_{Naiad}^4 \cos(73\lambda_{Thalassa} - 69\lambda_{Naiad} - 4\Omega_{Naiad}) \qquad (2)$$

$$\langle \mathcal{R} \rangle_2 = f_2 \, i_{Thalassa}^4 \cos(73\lambda_{Thalassa} - 69\lambda_{Naiad} - 4\Omega_{Thalassa}) \qquad (3)$$

$$\langle \mathcal{R} \rangle_3 = f_3 \, i_{Naiad}^2 i_{Thalassa}^2 \cos(73\lambda_{Thalassa} - 69\lambda_{Naiad} - 2\Omega_{Naiad} - 2\Omega_{Thalassa}) \qquad (4)$$

$$\langle \mathcal{R} \rangle_4 = f_4 \, i_{Naiad}^3 i_{Thalassa} \cos(73\lambda_{Thalassa} - 69\lambda_{Naiad} - 3\Omega_{Naiad} - \Omega_{Thalassa}) \qquad (5)$$

$$\langle \mathcal{R} \rangle_5 = f_5 \, i_{Naiad} i_{Thalassa}^3 \cos(73\lambda_{Thalassa} - 69\lambda_{Naiad} - \Omega_{Naiad} - 3\Omega_{Thalassa}) \qquad (6)$$

Here, $f_i = f_i(\alpha)$, are combinations of Laplace coefficients. We tested all five cases, and the resonant argument only librates for Eq. 2. The Naiad-Thalassa resonance also has five resonant terms equivalent to Eq. 2-6 which involve eccentricities, but these would have negligible contributions given the near-circular orbits of Naiad and Thalassa. There are also mixed inclination and eccentricity resonant terms that would also be very small.

The Naiad and Thalassa resonant argument can be converted from rates to periods for a more intuitive interpretation. First, we will adopt the following expression for the longitude of the node:

$$\dot{\Omega}_{Naiad} = \dot{\lambda}_{Naiad} - \dot{\nu}_{Naiad} \qquad (7)$$

Here, ν=M+ω is the vertical angle, M is the mean anomaly, and ω is the argument of periapsis.

$$73\dot{\lambda}_{Thalassa} - 69\dot{\lambda}_{Naiad} - 4(\dot{\lambda}_{Naiad} - \dot{\nu}_{Naiad}) = 0 \qquad (8)$$

$$73(\dot{\lambda}_{Thalassa} - \dot{\lambda}_{Naiad}) = -4\dot{\nu}_{Naiad} \qquad (9)$$

From Eq. 9, it is straightforward to replace the rates with periods as $P_{Naiad-Thalassa}^{synodic} = \frac{73}{4} P_{Naiad}^{vertical}$. This means that Naiad undergoes 18.25 vertical oscillations between every conjunction with Thalassa. Figure 5 illustrates what the fourth-order inclination resonance looks like in space in a frame that is centered on Neptune and that rotates with Thalassa. The large vertical amplitude is a consequence of Naiad's inclination with respect to the orbit of Thalassa. At the time that they align in mean longitude, Naiad is ~ 1850 km interior to Thalassa in the radial direction but $\frac{1}{\sqrt{2}} a_{Naiad} \sin(i_{Naiad}) \approx 2800\ km$ away vertically. From Thalassa's viewpoint, Naiad is ~57° out of the orbital plane at the conjunction point. This is a simplified description of integrated orbits. The sequence of symmetric offsets repeats every four synodic periods: south, south, north, north. Like many resonances, this particular configuration maximizes the minimum conjunction distance between the two moons.

The conjunction librates about the node of Naiad with an observed average amplitude of ~66° and a period of ~1.9 years. Figure 5 shows what the Naiad-Thalassa encounters look like when φ≈180° while Figure 6 shows resonant approaches when the libration amplitude is at its maximum. The inclination of Naiad with respect to Thalassa is no longer ~57° as it was in Figure 5, but it oscillates around this value from conjunction to conjunction. Again, this is a simplified description of integrated orbits.

While a fourth-order inclination resonance would normally be expected to be very weak, the large inclination of Naiad, combined with the small radial separation between the two moons, strengthens this interaction considerably. The high-inclination of Naiad is likely not caused by the fourth-order resonance with Thalassa, but by the past lower-order resonant captures. Banfield and Murray (1992) identified 35 second-order mean motion resonances, $\phi_{i_{Naiad}^2} = (p+2)\lambda_{satellite} - p\lambda_{Naiad} - 2\Omega_{Naiad}$, that Naiad could have passed through since Triton's orbit circularization. Banfield and Murray (1992) proposed that the high inclination most likely originates from a temporary 12:10 Naiad-Despina resonance that ended with the strengthening of the secondary 2:1 resonance. Banfield and Murray (1992) noted that the resonance order depends on the assumed masses of the satellites, but that the overall likelihood of a resonant capture does not. Another possible past resonance is the 2:1 Larissa-Naiad (Zhang and Hamilton, 2007). Naiad and Thalassa likely drifted into the 73:69 resonance under the tidal interactions with Neptune after the lower-order resonance kicked Naiad's inclination.

## 4.2 Accuracy of the orbital fits

The covariance matrix of parameter errors, produced by the weighted least-squares, is a measure of the fit accuracy. Jacobson et al. (2012) concluded that the orbital fit errors in general are dominated by random and systematic errors in observations as opposed to the orbital model being incomplete or the dynamical constants being poorly determined. The rms of the weighted residuals from Tables 1 and 2 suggest that the data were properly weighted and we did not notice any systematic biases when inspecting the residuals. We additionally inflated the uncertainties by including the $GM$ of the Neptune's system and the zonal harmonic $J_4$ in the consider parameter list. It is thus reasonable to assume that the formal errors from the fit are good approximation of the orbital uncertainties. We considered including Triton's orbit into this fit as a potential source of unmodeled uncertainties, but Jacobson (2009) showed that the Triton orbit uncertainty in the in-orbit direction grows linearly to only 75 km (~3.7 mas) for the period 1990-2020. This is below the current data accuracy and should not affect the fit.

The orbital uncertainties are often used to evaluate the need for future observations. We used linear covariance mapping as described in Jacobson et al. (2012) to propagate the orbital uncertainties to 2020-2030. Figure 7 shows the combined in-orbit (tangential), radial, and out-of-plane (normal) positional uncertainties. All uncertainties are on the order of several tens to several hundred km, which means that the satellites are in no danger of being lost. The largest uncertainty component is in-orbit direction due to the uncertainties in the orbital mean motion. The uncertainty in Neptune's pole orientation coupled with the uncertainty in the Neptune's gravitational harmonics are the primary causes of the out-of-plane uncertainties. It is important to note that there is a definite need for more astrometry of the regular moons of Neptune, in this case not due to large orbital uncertainties, but due to an interesting resonant dynamics that may lead to better *GM* constraints for Naiad, Thalassa, and Proteus.

**4.3 Influence of Triton**

Figures 8A and 8B show the comparison between a precessing ellipse and an integrated orbit for Proteus over intervals of twenty and nine hundred years, respectively. These figures visualize orbital dynamics captured in the integration, but not in an analytical fit. The oscillations of several km present in Figure 8A in in-orbit direction have a period of ~12.8 years, or the same as the nodal precession of Proteus in Table 7**.** Intuitively, this can be explained by the perturbations from Triton that are part of the integrated orbit, but not part of the analytical solution. The mean motion of Proteus depends on the radial force; Neptune pulls inward while Triton pulls outward. At the moment when the ascending node of Proteus is aligned with Triton's descending node, the mean outward pull from Triton is at its strongest because the orbits are closer to the same plane. This is the point when Proteus slows down. The opposite occurs half a mutual nodal regression period later. The orbital planes are now most distant from each other which means that the mean outward force of Triton on Proteus is a bit smaller and the moon speeds up. Figure 8B shows the longer-term variations in the orbit of Proteus due to the nodal precession of Triton which is ~687 years long (Jacobson, 2009). The same trend is visible for other satellites, although the magnitude decreases as the distance from Triton increases.

5. Conclusions

We report on the latest orbital fits for the regular satellites of Neptune. This analysis is an extension of the Jacobson and Owen (2004) and Jacobson (2009) orbital fits, but with a much more extensive set of astrometry, and with a more sophisticated orbital model that uses numerically integrated equations of motion. We were able to estimate the *GMs* for Naiad and Thalassa, $GM_{Naiad}$= 0.0080±0.0043 km$^3$ s$^{-2}$ and $GM_{Thalassa}$=0.0236±0.0064 km$^3$ s$^{-2}$, due to the fact that the satellites are in a 73:69 inclination resonance. This fourth-order resonance is unique among the planetary satellites. The resonant argument librates with a large amplitude, ~66°, and a fast period, ~1.9 years, when compared to the values known for first- and second- order mean motion resonances that involve planets or satellites in the solar system (Table 8.8 in Murray and Dermott, 2001). However, these values were calculated based on the nominal set of masses which may change with the addition of more high-precision astrometry. We also found that the newly discovered moon Hippocamp is in a 13:11 near-resonance with Proteus. This is only the third second-order resonance among outer planet moons after the 4:2 Mimas-Tethys and the 77:75 Methone-Anthe. Future observations of Hippocamp could reveal the mass of Proteus. The *GM*s of Despina, Galatea, and Larissa are more difficult to measure because they are not in any direct resonance and their masses are small. Our fit also yielded an improved value for the oblateness coefficient of Neptune: $J_2$=3409.1±2.9 × 10$^{-6}$. The orbital uncertainties show that the positions of the satellites are known within several hundred kilometers until at least 2030.

## Acknowledgements

The work of MB and RAJ was performed at the Jet Propulsion Laboratory, California Institute of Technology, under contract with the National Aeronautics and Space Administration (NASA). MRS and RSF were supported in part by NASA's Outer Planets program through grant NNX14AO40G and by the Space Telescope Science Institute through program HST-GO-14217. The authors would like to thank the two anonymous reviewers for their comments which improved this manuscript.

# Tables

**Table 1.** Voyager 2 observations.

| Satellite | Time | Observatory | Points | Astrometry Type | Reference | Residuals (km, km) | Nresid. |
|---|---|---|---|---|---|---|---|
| Naiad    | 1989 | Voyager 2 | 28,28   | Sample, Line | Owen et al. (1991), Jacobson and Owen (2004) | 41, 27  | 0.81,0.58 |
| Thalassa |      |           | 45,45   |              |                                              | 45, 40  | 0.48,0.48 |
| Despina  |      |           | 104,104 |              |                                              | 62, 68  | 0.54,0.52 |
| Galatea  |      |           | 101,101 |              |                                              | 91, 62  | 0.72,0.48 |
| Larissa  |      |           | 138,138 |              |                                              | 88, 66  | 0.61,0.45 |
| Proteus  |      |           | 178,178 |              |                                              | 155,160 | 0.54,0.54 |

Samples and lines determine locations of satellites in Voyager 2 images. The last two columns list the residuals for the nep090 orbital fit and the same residuals normalized by the data weights (Nresid).

**Table 2.** The Earth-based and the HST observations.

| Satellite | Time | Observatory | Points | Astrometry Type | Reference | Residuals (mas, mas) | Nresid. |
|---|---|---|---|---|---|---|---|
| Larissa  | 1981 | Mt Lemon      | 1,1   | RA, dec         | Reitsema et al. (1982), | 15,185   | 0.08,0.93 |
|          | 1981 | Catalina      | 1,1   |                 | Jacobson and Owen (2004) | 15,185  | 0.08,0.93 |
| Proteus  | 1991 | La Silla      | 8,8   | X, Y            | Colas and Buil (1992)   | 97,124   | 0.97,0.99 |
| Despina  |      | HST           | 6,6   | $\theta, \rho$  | Pascu et al. (2004)     | 15, 18   | 0.95,1.11 |
| Galatea  |      |               | 17,17 |                 |                         | 18, 16   | 1.07,0.94 |
| Larissa  |      |               | 13,5  |                 |                         | 11,  9   | 1.07,0.93 |
| Proteus  |      |               | 39,39 |                 |                         |  7,  6   | 0.98,0.86 |
| Galatea  | 1998 | Mauna Kea     | 8,8   | X, Y            | Sicardy et al. (1999)   | 67, 45   | 0.96,1.00 |
| Despina  | 1998 | HST           | 1     | $\Delta_{long}$ | Dumas et al. (2002)     | 15       | 0.29 |
| Galatea  |      |               | 1     |                 |                         | 17       | 0.55 |
| Larissa  |      |               | 1     |                 |                         | 14       | 0.92 |
| Proteus  |      |               | 1     |                 |                         |  0       | 0.03 |
| Proteus  | 2000 | Picos Dos Dias| 4,4   | X, Y            | Vieira Martins          | 243, 83  | 1.06,0.33 |
| Proteus  | 2001 |               | 16,16 |                 | et al. (2004)           | 229, 82  | 1.00,0.33 |
| Proteus  | 2002 |               | 42,42 |                 |                         | 150, 91  | 0.65,0.36 |
| Thalassa | 2002 | Mauna Kea     | 2,2   | X, Y            | Marchis et al. (2004)   | 450,272  | 1.80,1.36 |
| Despina  |      |               | 9,9   |                 |                         | 36, 32   | 1.28,1.29 |
| Galatea  |      |               | 15,15 |                 |                         | 37, 31   | 1.34,1.23 |
| Larissa  |      |               | 15,15 |                 |                         | 46, 27   | 1.12,1.08 |
| Proteus  |      |               | 16,16 |                 |                         | 35, 26   | 1.09,1.05 |
| Despina  | 2003 | Mauna Kea     | 16,16 | X, Y            | Marchis et al. (2004)   | 16, 11   | 0.56,0.45 |
| Galatea  |      |               | 18,18 |                 |                         | 16, 11   | 0.57,0.45 |
| Larissa  |      |               | 20,20 |                 |                         | 15, 11   | 0.53,0.44 |
| Proteus  |      |               | 12,12 |                 |                         |  8,  8   | 0.27,0.31 |
| Naiad    | 2004 | HST           | 1,1   | X, Y            | Showalter et al. (2019) | 11,  4   | 1.68,0.61 |
| Thalassa |      |               | 7,7   |                 |                         |  7, 13   | 1.09,1.88 |
| Despina  |      |               | 25,25 |                 |                         |  6,  4   | 1.24,0.86 |
| Galatea  |      |               | 42,42 |                 |                         |  5,  4   | 0.99,0.80 |
| Larissa  |      |               | 35,35 |                 |                         |  2,  2   | 0.67,0.75 |
| Proteus  |      |               | 50,50 |                 |                         |  3,  2   | 1.06,0.73 |
| Hippocamp|      |               | 3,3   |                 |                         |  3,  3   | 0.69,0.80 |
| Thalassa | 2005 | HST           | 7,7   | X, Y            | Showalter et al. (2019) |  7, 10   | 1.51,1.30 |
| Despina  |      |               | 64,64 |                 |                         |  4,  4   | 0.91,0.92 |
| Galatea  |      |               | 87,87 |                 |                         |  3,  4   | 0.85,0.96 |
| Larissa  |      |               | 102,102 |               |                         |  4,  3   | 1.12,0.94 |
| Proteus  |      |               | 103,103 |               |                         |  2,  2   | 0.77,0.84 |
| Hippocamp|      |               | 1,1   |                 |                         |  5, 15   | 0.76,2.17 |
| Naiad    | 2009 | HST | 5,5   | X, Y | Showalter et al. (2019) | 23, 13 | 1.53,0.91 |
| Thalassa |      |     | 3,3   |      |                         | 10,  5 | 1.25,0.85 |
| Despina  |      |     | 19,19 |      |                         |  5,  4 | 0.98,1.02 |
| Galatea  |      |     | 40,40 |      |                         |  5,  5 | 0.78,0.75 |
| Larissa  |      |     | 30,30 |      |                         |  4,  3 | 0.77,0.61 |
| Proteus  |      |     | 44,44 |      |                         |  2,  2 | 0.62,0.49 |
| Hippocamp|      |     | 3,3   |      |                         |  1, 11 | 0.24,1.13 |
| Despina  | 2010 | HST | 23,23 | X, Y | Showalter et al. (2019) |  9,  7 | 1.14,0.87 |
| Galatea  |      |     | 35,35 |      |                         |  9,  6 | 1.02,0.79 |
| Larissa  |      |     | 29,29 |      |                         |  7,  6 | 1.17,0.98 |
| Proteus  |      |     | 36,36 |      |                         |  2,  3 | 0.54,0.69 |
| Despina  | 2011 | HST | 37,37 | X, Y | Showalter et al. (2019) |  9,  9 | 0.91,0.96 |
| Galatea  |      |     | 24,24 |      |                         |  7,  8 | 0.76,0.88 |
| Larissa  |      |     | 49,49 |      |                         |  8,  7 | 1.02,0.91 |
| Proteus  |      |     | 63,63 |      |                         |  3,  4 | 0.74,0.88 |
| Despina  | 2015 | HST | 32,32 | X, Y | Showalter et al. (2019) | 25, 18 | 1.02, 0.84 |
| Galatea  |      |     | 58,58 |      |                         | 22, 15 | 1.10, 0.94 |
| Larissa  |      |     | 56,56 |      |                         | 16, 14 | 0.94, 1.00 |
| Proteus  |      |     | 57,57 |      |                         |  5,  5 | 1.14, 1.23 |
| Naiad    | 2016 | HST | 10,10 | X, Y | Showalter et al. (2019) | 15, 11 | 1.40, 1.19 |
| Thalassa |      |     | 29,29 |      |                         | 11, 12 | 1.34, 1.22 |
| Despina  |      |     | 25,25 |      |                         |  3,  3 | 0.71, 0.84 |
| Galatea  |      |     | 28,28 |      |                         |  4,  4 | 0.88, 0.96 |
| Larissa  |      |     | 35,35 |      |                         |  6,  2 | 0.80, 0.43 |
| Proteus  |      |     | 43,43 |      |                         |  3,  1 | 0.69, 0.34 |
| Hippocamp|      |     | 9,9   |      |                         |  6,  5 | 0.82, 0.86 |

Astrometry in the format X, Y is relative separations of a satellite and Neptune on the tangent plane in RA and Dec direction. θ, ρ are satellite's position angle and separation from Neptune, $\Delta_{long}$ is the satellite's longitudinal separation from Neptune. RA, Dec are absolute measurements of the satellite position on the plane of sky. The last two columns list the residuals for the nep090 orbital fit and the same residuals normalized by the data weights (Nresid).

**Table 3.** Dynamical constants used in the orbital fit

| Parameter | Value | Reference |
|---|---|---|
| $GM$ Neptune system | $6836527.1 \pm 10.0$ km$^3$s$^{-2}$ | Jacobson (2009) |
| $GM$ Triton | $1427.6 \pm 1.9$ km$^3$s$^{-2}$ | Jacobson (2009) |
| $GM$ Jupiter system | $126712764.1 \pm 2.7$ km$^3$s$^{-2}$ | JUP310 Jacobson, personal comm. |
| $GM$ Saturn system | $37940584.9 \pm 0.04$ km$^3$s$^{-2}$ | Jacobson, personal comm. |
| $GM$ Uranus system | $5794556.5 \pm 4.3$ km$^3$s$^{-2}$ | Jacobson (2014) |
| $GM$ Sun+ | $132713233264. \pm 10.$ km$^3$s$^{-2}$ | Konopliv (2011); Folkner (2016) |
| $J_4$ Neptune ($\times 10^{-6}$) | $-33.4 \pm 2.9$ | Jacobson (2009) |

Parameters and their values that were used in nep090 orbital solution. *GM* Sun+ stands for the mass of the Sun augmented by the masses of the terrestrial planets and the Moon. $J_2$ is estimated in the fit, while $J_4$ value is held fixed based on Voyager 2 data (Jacobson, 2009). JUP310 (Jacobson, personal comm.) are ephemerides for the Galilean satellites available on JPL Horizons database: https://ssd.jpl.nasa.gov/horizons.cgi (Giorgini et al., 1996).

**Table 4.** State vectors and their uncertainties.

| Satellite | Position (km) | σ (km) | Velocity (km s$^{-1}$) | σ (km s$^{-1}$) |
|---|---|---|---|---|
| Naiad     | 38622.15178849510  | 20.  | -5.192078367667181 | 0.00704 |
|           | 28591.24153864961  | 21.  |  5.456862658299498 | 0.00621 |
|           |  4864.059370324697 | 31.  |  9.226693664145152 | 0.00254 |
| Thalassa  | 42788.42603660395  | 14.  | -4.386937928695694 | 0.00482 |
|           | 26055.26565945493  | 21.  |  6.649099992784474 | 0.00425 |
|           |  1739.911923856407 | 22.  |  8.558823312442090 | 0.00331 |
| Despina   | -32183.30540272224 | 12.  |  8.034862203275067 | 0.00171 |
|           | -37295.00414520138 |  9.  | -3.375316024267507 | 0.00253 |
|           | -17981.08043273324 | 10.  | -7.377004848585427 | 0.00150 |
| Galatea   | -25919.69475269366 | 13.  |  8.767539491626424 | 0.00113 |
|           | -47166.08851829967 |  6.  | -1.143027297382970 | 0.00190 |
|           | -30518.35663600751 | 10.  | -5.684044107376814 | 0.00145 |
| Larissa   |  58494.36328593099 |  7.  |  4.732561367834237 | 0.00139 |
|           | -12639.69688416622 | 12.  |  7.202403080805486 | 0.00086 |
|           | -42901.10080049761 | 11.  |  4.303239131788079 | 0.00124 |
| Hippocamp |  85048.87057480606 | 149. | -3.762159902153145 | 0.01783 |
|           |  60927.21769143848 | 177. |  4.070644538587873 | 0.01532 |
|           |  12429.89739095648 | 237. |  5.847629561228800 | 0.00638 |
| Proteus   | -46301.98049762944 | 12.  |  6.445686731041889 | 0.00038 |
|           | -89449.00901759176 |  5.  | -0.6112611011524898 | 0.00067 |
|           | -60730.36015174328 |  8.  | -4.020949170553930 | 0.00047 |

The state vectors (nep090) are listed at epoch 1989 September 01, 00:00:00 TDB (barycentric dynamical time). We list the full precision state vectors suitable for numerical integration and the formal 1σ uncertainties from the fit.

**Table 5.** Masses and densities of the inner moons of Neptune

| Satellite | Radii (km) | Mass* x10$^{15}$ (kg) | GM* (km$^3$s$^{-2}$) | Fitted GM (km$^3$s$^{-2}$) | GM uncertainty (km$^3$s$^{-2}$) | Density (g cm$^{-3}$) |
|---|---|---|---|---|---|---|
| Naiad     | 33±3  | 7.527–225.8   | 0.0005–0.0151 | 0.0080         | 0.0043 | 0.80±0.48 |
| Thalassa  | 41±3  | 14.43–433.0   | 0.0010–0.0289 | 0.0236         | 0.0064 | 1.23±0.43 |
| Despina   | 75±3  | 88.36–2651.   | 0.0059–0.1769 | 0.1179$^{Ap}$  | 0.2428 | – |
| Galatea   | 88±4  | 142.7–4282.   | 0.0095–0.2857 | 0.1905$^{Ap}$  | 0.9816 | – |
| Larissa   | 97±3  | 191.1–5734.   | 0.0128–0.3827 | 0.2551$^{Ap}$  | 4.8479 | – |
| Hippocamp | 17±2  | 1.029–30.87   | 0.0001–0.0021 | 0.0014$^{Ap}$  | 0.8006 | – |
| Proteus   | 210±7 | 1940.–58190.  | 0.1294–3.8829 | 2.5886$^{Ap}$  | 4.6455 | – |

Radii estimates for all but Hippocamp are based on Voyager 2 imaging data (Karkoschka, 2003). Hippocamp radius is from Showalter et al. (2019). The masses and *GM*s marked with an asterisk are calculated by assuming densities of 0.05–1.5 g cm$^{-3}$. Here, *GM* is the product of the Newtonian constant of gravitation, $G = 6.67300 \times 10^{-11}$ m$^3$ kg$^{-1}$ s$^{-2}$, and the body's mass, *M*. The *GM*s marked with Ap had an *a priori GM* set to the density of 1.0±0.2 g cm$^{-3}$. The *GM* uncertainties represent the formal post-fit 1σ without any *a priori* values. The densities were calculated only for Naiad and Thalassa because their masses were fit without any a priori values. The uncertainties were obtained by propagating errors on radius and mass.

**Table 6.** Comparison of the massless and nep090 solutions for Naiad and Thalassa

|  | Naiad | | | | Thalassa | | | |
|---|---|---|---|---|---|---|---|---|
|  | Voyager 2 | | Showalter et al. (2019) | | Voyager 2 | | Showalter et al. (2019) | |
|  | Sample (km) | Line (km) | X (mas) | Y (mas) | Sample (km) | Line (km) | X (mas) | Y (mas) |
| Massless | 3.0±44.5 | 7.3±27.6 | −10.0±17.0 | −1.0±12.0 | −12.8±45.0 | −7.7±39.1 | 2.0±10.0 | 0.0±12.0 |
| nep090 | 2.7±40.6 | 6.6±27.0 | −10.0±18.0 | −1.0±11.0 | −11.0±44.6 | −6.1±39.5 | 2.0±10.0 | 0.0±12.0 |

List of means and rms of the residuals for two different orbital fits. The first orbital solution assumed that the satellites are massless. The second solution is our nominal nep090 fit where the masses were allowed to adjust. The residuals are shown separately for Voyager 2 and Showalter et al. (2019) data.

**Table 7.** Neptune-centric mean orbital elements at 2020 January 1 TDB referred to the local Laplace planes

| Element | Naiad | Thalassa | Despina | Galatea | Larissa | Hippocamp | Proteus |
|---|---|---|---|---|---|---|---|
| a (km) | 48228. | 50075. | 52526. | 61953. | 73548. | 105283. | 117647. |
| e | 0.00014 | 0.00019 | 0.00027 | 0.00020 | 0.00121 | 0.00001 | 0.00047 |
| $\varpi(°)$ | 194.476 | 66.367 | 320.236 | 137.266 | 191.273 | 55.913 | 301.616 |
| $\omega(°)$ | 295.867 | 42.832 | 215.729 | 258.396 | 234.488 | 305.446 | 268.929 |
| $\lambda(°)$ | 98.810 | 341.813 | 197.617 | 107.056 | 72.262 | 25.814 | 70.716 |
| M(°) | 264.335 | 275.446 | 237.381 | 329.789 | 240.988 | 329.901 | 129.101 |
| $i_{free}(°)$ | 4.728 | 0.168 | 0.039 | 0.010 | 0.214 | 0.002 | 0.042 |
| $\Omega(°)$ | 258.609 | 23.535 | 104.507 | 238.870 | 316.786 | 110.467 | 32.686 |
| $\dot\lambda(° \text{ day}^{-1})$ | 1222.858 | 1155.759 | 1075.733 | 839.661 | 649.054 | 378.906 | 320.766 |
| $\dot\varpi(° \text{ yr}^{-1})$ | 620.264 | 551.560 | 466.078 | 261.265 | 143.320 | 37.534 | 28.147 |
| $\dot\omega(° \text{ yr}^{-1})$ | 1246.149 | 1102.393 | 931.603 | 522.308 | 286.555 | 77.358 | 56.289 |
| $\dot\Omega(° \text{ yr}^{-1})$ | -625.885 | -550.833 | -465.524 | -261.043 | -143.236 | -39.824 | -28.142 |
| $P_\lambda$(days) | 0.29 | 0.31 | 0.33 | 0.43 | 0.55 | 0.95 | 1.12 |
| $P_\varpi$(yr) | 0.58 | 0.65 | 0.77 | 1.38 | 2.51 | 9.59 | 12.79 |
| $P_\omega$(yr) | 0.29 | 0.33 | 0.39 | 0.69 | 1.26 | 4.65 | 6.40 |
| $P_\Omega$(yr) | 0.58 | 0.65 | 0.77 | 1.38 | 2.51 | 9.04 | 12.79 |
| $i_{Lap}(°)$ | 0.467 | 0.468 | 0.470 | 0.481 | 0.510 | 0.773 | 1.010 |
| $\alpha_{Lap}(°)$ | 299.459 | 299.459 | 299.459 | 299.459 | 299.458 | 299.457 | 299.456 |
| $\delta_{Lap}(°)$ | 42.938 | 42.937 | 42.935 | 42.923 | 42.895 | 42.632 | 42.395 |

Elements include semi-major axis, a, eccentricity, e, inclination $i_{Lap}$ of the local Laplace plane to the invariable plane, inclination $i_{free}$ of the orbit plane to the local Laplace plane, mean longitude $\lambda$, mean anomaly M, and longitude of periapsis, $\varpi = \Omega+\omega$, where $\omega$ is argument of periapsis and $\Omega$ is the longitude of the node. The longitudes are measured from the intersection of the local Laplace planes with the invariable reference plane so these are the equivalent of "free" angles in Zhang and Hamilton (2007). The nodal and longitudinal rates are corrected for the local Laplace plane precession about the invariable plane. The pole of the invariable plane is $\alpha_{inv}$=299.4609°, $\delta_{inv}$=43.4048° (Jacobson, 2009).

**Figures**

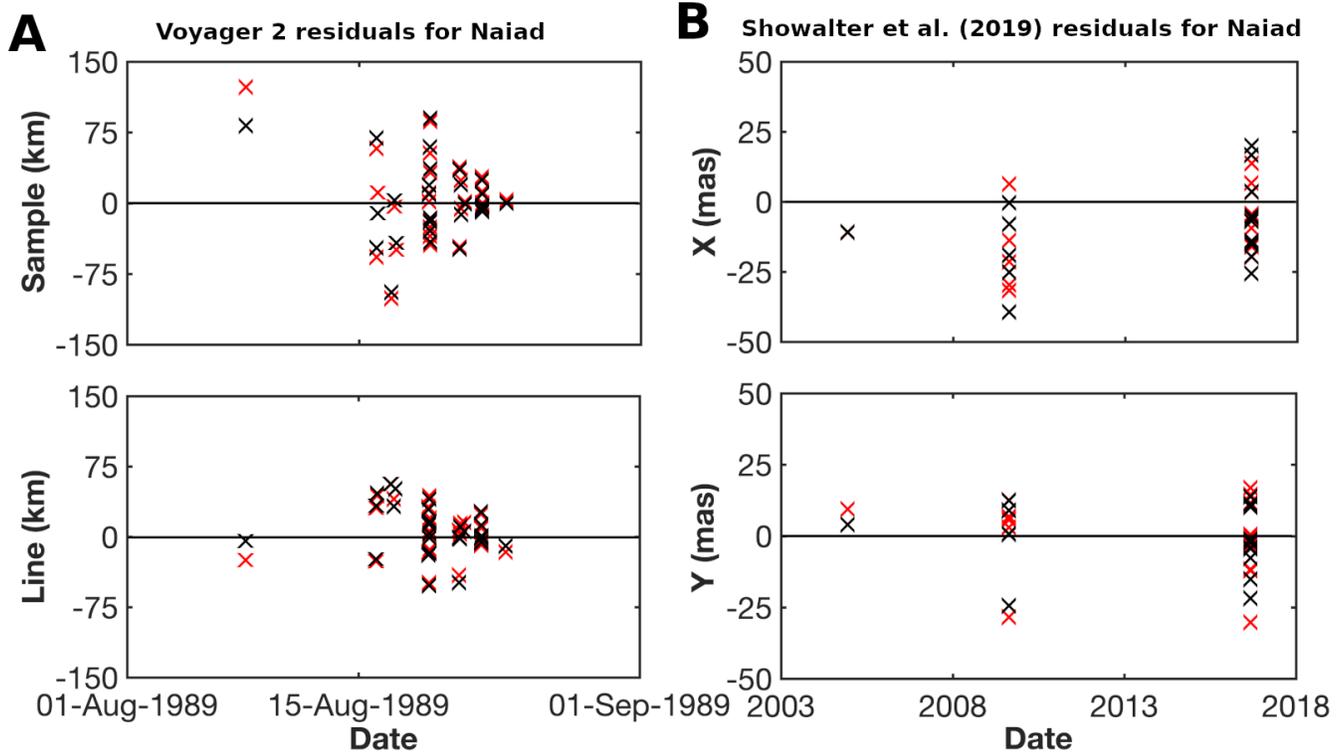

**Figure 1. A.** Naiad residuals for the integrated orbit fit for the 28 astrometry points from Voyager 2. The red points mark the residuals for an orbital solution where the masses of the satellites are zero. The black points mark the residuals for nep090 solution where the satellites have mass. **B.** Naiad residuals for the 16 astrometry points from Showalter et al. (2019).

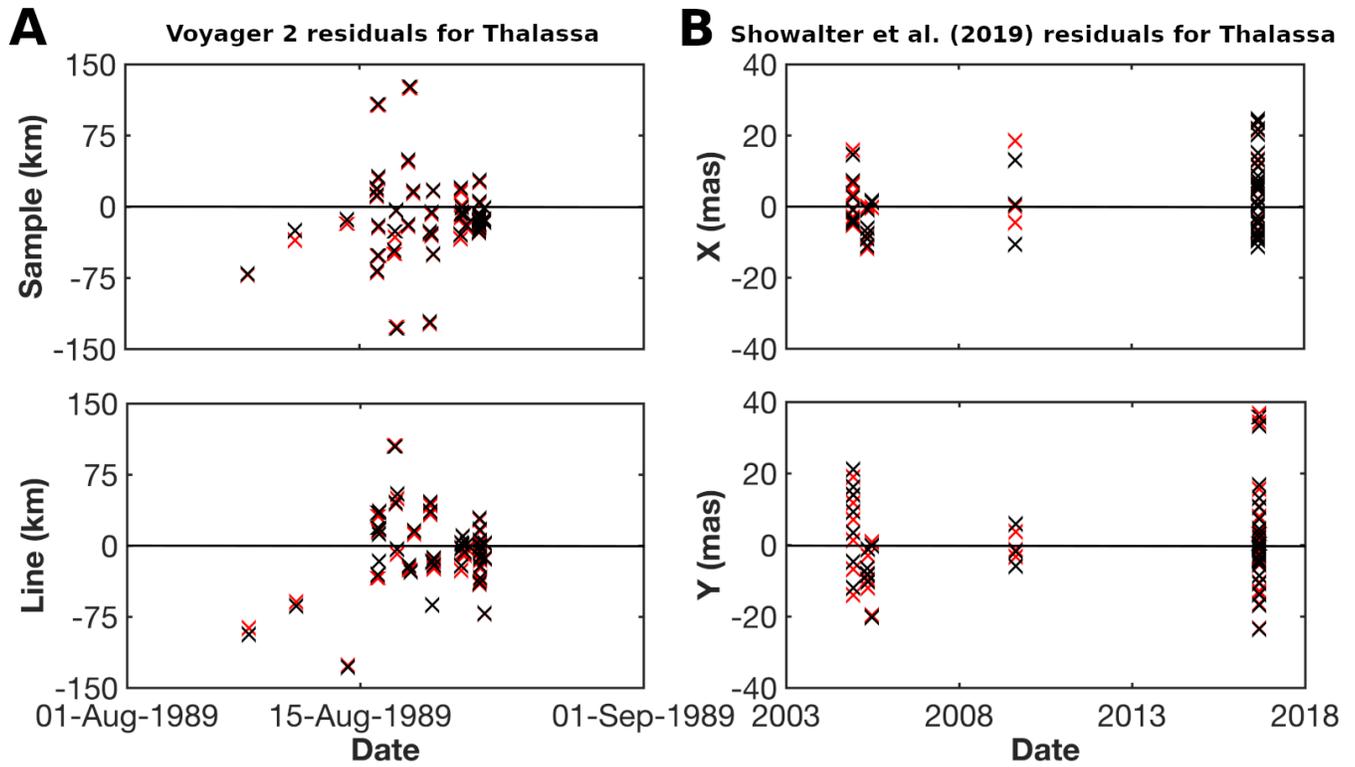

**Figure 2. A.** Thalassa residuals for the integrated orbit fit for the 45 astrometry points from Voyager 2. The red points mark the residuals for an orbital solution where the masses of the satellites are zero. The black points mark the residuals for nep090 solution where the satellites have mass. **B.** Thalassa residuals for the 46 astrometry points from Showalter et al. (2019).

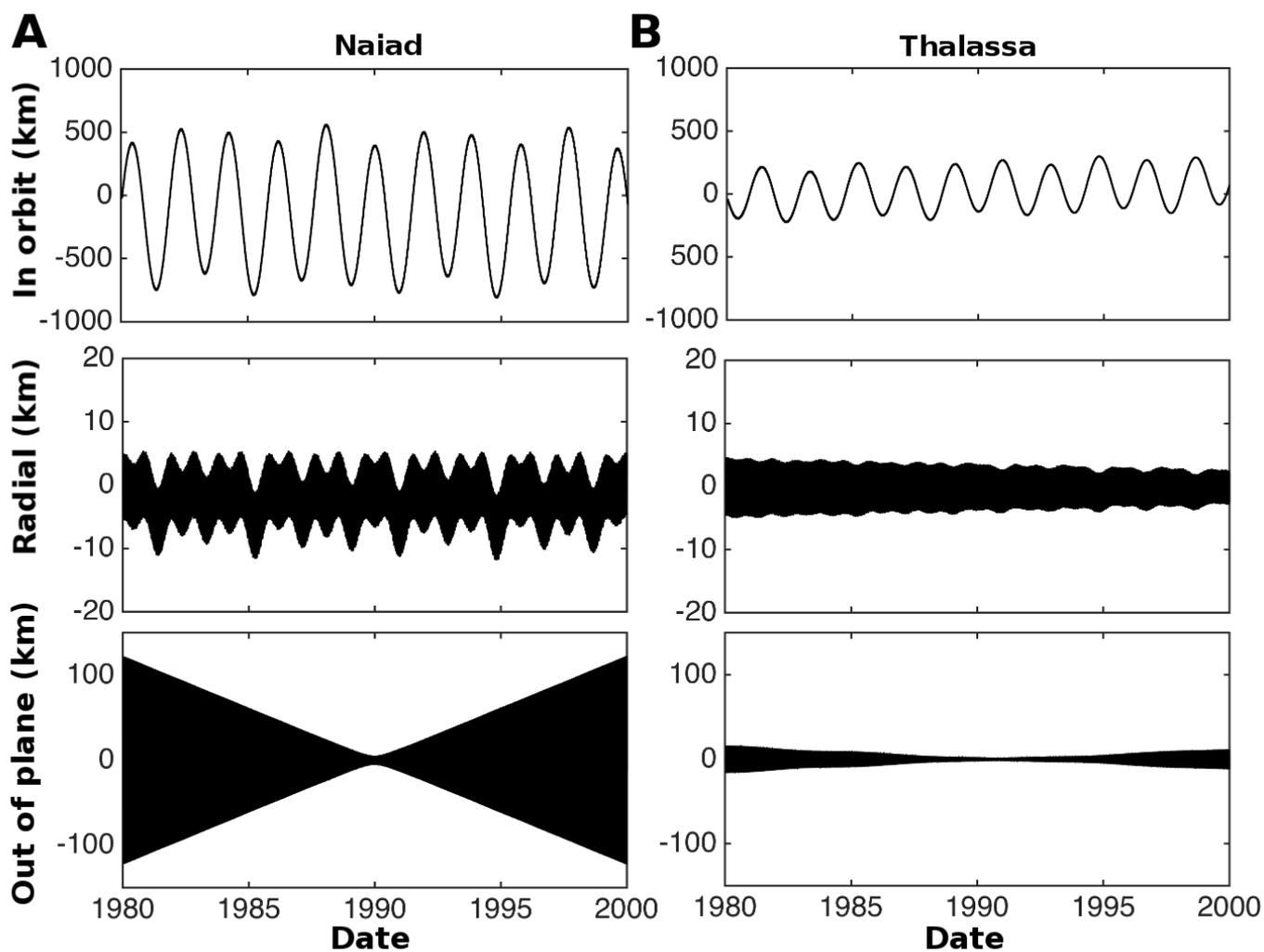

**Figure 3.** Comparison between the massless and with-mass (nep090) orbital solutions for Naiad and Thalassa. **A.** In-orbit (along the track), radial, and out-of-plane differences for the orbit of Naiad. **B.** In-orbit, radial, and out-of-plane differences for the orbit of Thalassa.

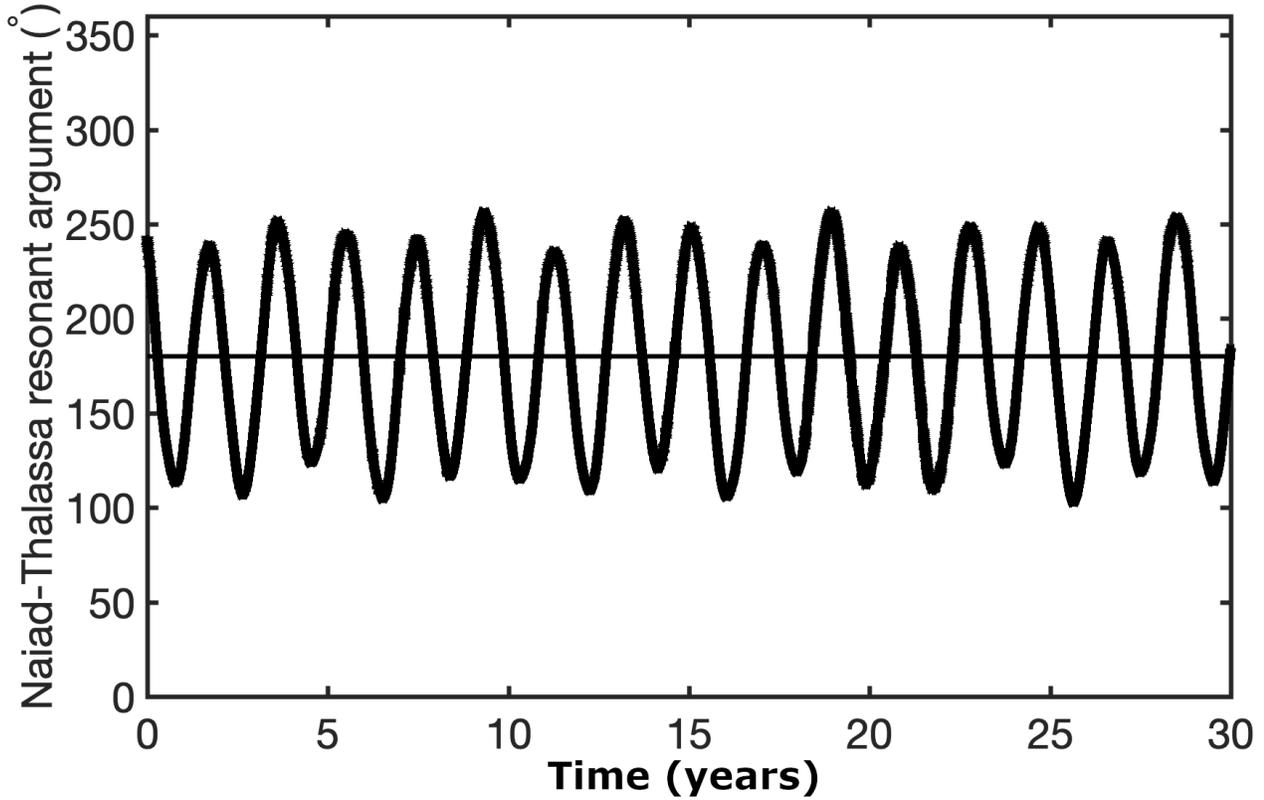

**Figure 4.** The resonant angle $\varphi = 73\lambda_{Thalassa} - 69\lambda_{Naiad} - 4\Omega_{Naiad}$ calculated based on osculating elements from integrated orbits (nep090). We used a function, $a_0 + a_1 \cos(\omega_1 x) + a_2 \sin(\omega_1 x)$, where x is time in years, in order to estimate an amplitude and a period of libration. The fit yielded $a_0 = 179.9°$, $a_1 = 52.1°$, $a_2 = -40.7°$, and $P = \frac{2\pi}{\omega_1} = 1.9\ years$. The average amplitude of libration is ~66°. The dates and resonant angle values for one full libration cycle are: 2018 Apr. 20, 16:48 UTC – 2018 Sep. 27, 16:48 UTC – 2019 Mar. 12, 12:00 UTC – 2019 Sep. 19, 12:00 UTC – 2020 Mar. 09, 19:12 UTC and 180.0°–235.8°–180°–108.7°–180° respectively.

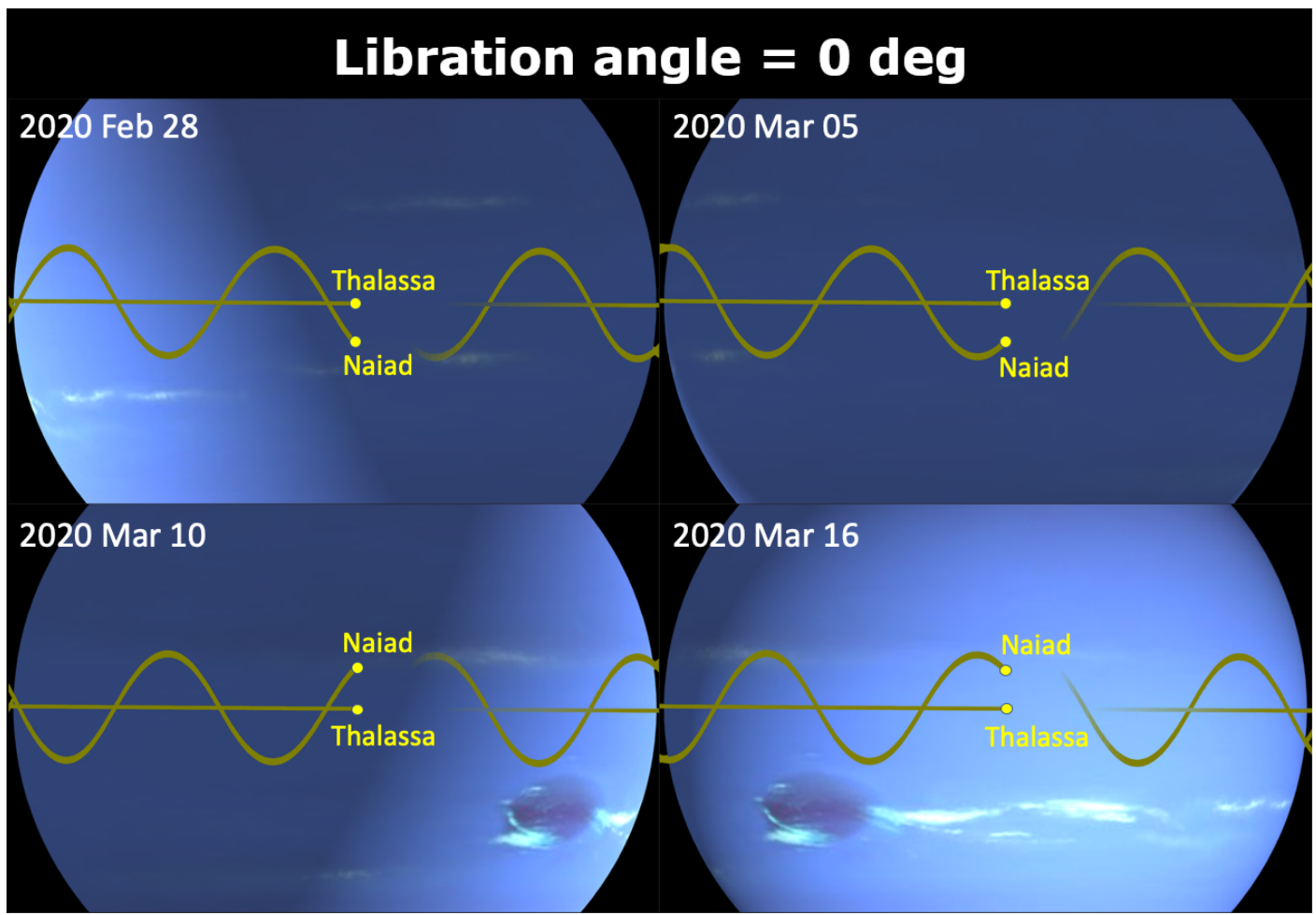

**Figure 5.** Naiad-Thalassa fourth-order mean motion inclination resonance. The orbit of Naiad is shown in a frame that is centered on Neptune and that rotates with Thalassa. The panels show four consecutive conjuctions of Naiad and Thalassa. Naiad undergoes 18.25 vertical oscillations from conjuction to conjuction which results in south, south, north, north offsets. The cycle repeats after this. At the time of a conjuction, Naiad is ~1850 km interior to Thalassa in the radial direction and ~2800 km in the vertical direction. This is a simplified description of integrated orbits. We used the data from Figure 4 to find the time when osculating elements add up to φ=180° and the libration angle is 0°. The visualization of Naiad and Thalassa integrated orbits (nep090) was done using the Cosmographia software, https://naif.jpl.nasa.gov/naif/cosmographia.html.

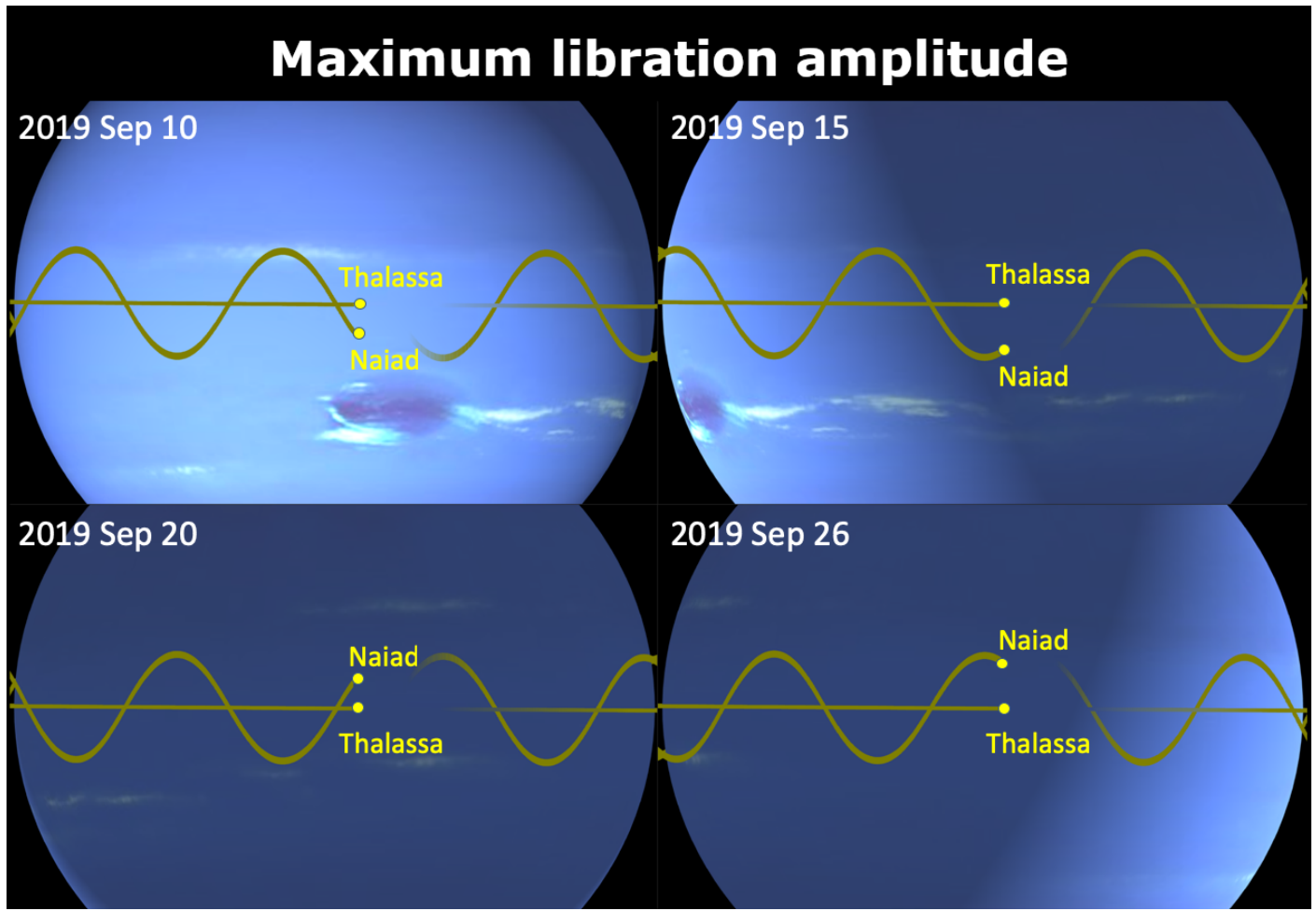

**Figure 6.** Naiad-Thalassa resonant encounters at the time of maximum libration angle. We used the data from Figure 4 to find the maximum libration amplitude. The rest of the caption is equivalent to Figure 5.

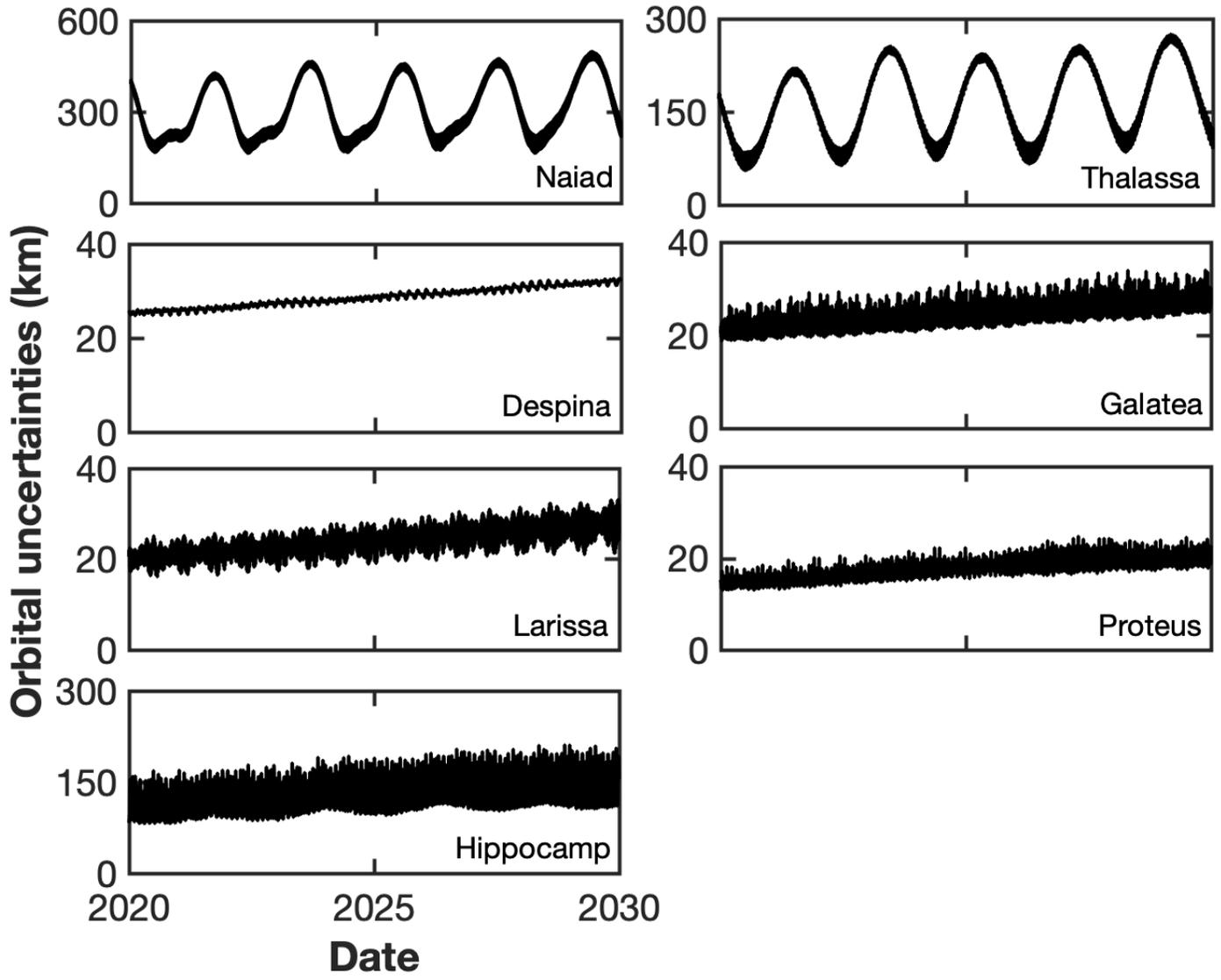

**Figure 7.** Estimated orbital uncertainties, $\sigma = \sqrt{\sigma_{in\ orbit}^2 + \sigma_{radial}^2 + \sigma_{out\ of\ plane}^2}$ for the orbital solution nep090.

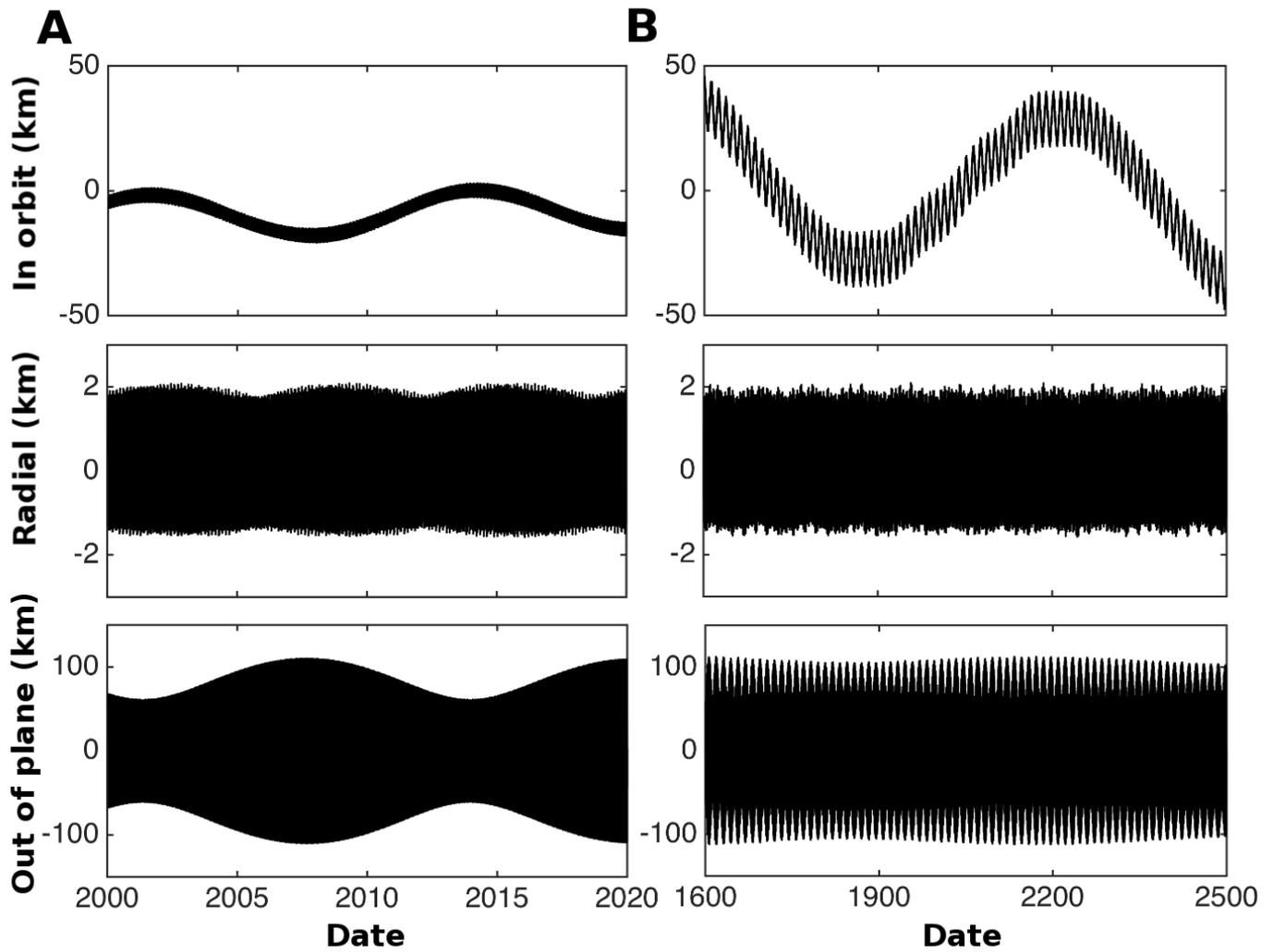

**Figure 8.** In-orbit, radial, and out-of-plane differences between an integrated orbit and a precessing ellipse approximation for the orbit of Proteus. **A.** The differences over twenty years, 2000-2020. **B.** The differences over nine hundred years, 1600-2500.